\documentclass[aps,prb,twocolumn,superscriptaddress,showpacs]{revtex4}
\usepackage{color}
\usepackage{amsmath}
\usepackage{verbatim}
\usepackage{hyperref}
\usepackage{epsf}
\usepackage{graphics}
\usepackage{graphicx}
\usepackage{subfigure}
\usepackage{natbib}
\usepackage{dcolumn}
\usepackage{bm}

\newcommand{\ket}[1]{|#1\rangle}

\begin{document}
\title{Magnetic breakdown in ortho-II high-temperature cuprates}
\author{Jean-Michel Carter}
\affiliation{Department of Physics, University of Toronto, Toronto, Ontario M5S 1A7 Canada}
\author{Daniel Podolsky}
\affiliation{Department of Physics, University of Toronto, Toronto, Ontario M5S 1A7 Canada}
\affiliation{Physics Department, Technion, Haifa 32000, Israel}
\author{Hae-Young Kee}
\email{hykee@physics.utoronto.ca}
\affiliation{Department of Physics, University of Toronto, Toronto, Ontario M5S 1A7 Canada}
\affiliation{School of Physics, Korea Institute for Advanced Study, Seoul 130-722, Republic of Korea}
\begin{abstract}

The recent confirmation of closed Fermi surface in the high temperature cuprates via de Haas-van Alphen and Shubnikov-de haas oscillations calls for a theoretical investigation of the origin of such oscillations. We study de Haas-van Alphen oscillations and magnetic breakdown in the context of ortho-II high temperature cuprates to understand the origin of the multiple frequencies. We find that the magnetic breakdown is highly sensitive not only to the ortho-II potential, but also to the Fermi surface topology, and is thus useful to distinguish between various theoretical proposals related to quantum oscillations.

\end{abstract}

\maketitle

\section{Introduction}

A remarkable phenomenon associated with a Fermi surface in a metal
is the observation of periodic oscillations in thermodynamic and transport quantities as a function
of the inverse of the magnetic field. This quantum effect is due to a quantization of 
Landau levels,  and its periodicity is proportional to the area enclosed by the Fermi surface in
two dimensions (or the area enclosed by extremal orbits in three dimensions).
One of the unsolved puzzles in the high temperature (high T$_c$) cuprates is a truncated Fermi surface,
dubbed the Fermi arc, detected by angle resolved photoemission spectroscopy (ARPES).\cite{Damascelli, Shen, Norman, Arpes} This cannot be understood within a conventional theory -- for instance, if the truncation
occurred due to a broken translational symmetry, this would lead to Brillouin zone folding,
which has not been observed so far in ARPES.\cite{Zhou}  This unconventional metallic state
with a Fermi arc has been called the pseudogap phase.
Up until 2007, it seemed that quantum oscillations failed to occur in the
pseudogap state of high T$_c$ cuprates.

Since the first striking report of the high T$_c$ cuprates by Doiron-Leyraud {\em et al.} \cite{Taillefer} in 2007,
a series of experiments have reported quantum oscillations in
YBa$_2$Cu$_3$O$_{6.51}$(YBCO$_{6.51}$)  and YBa$_2$Cu$_4$O$_8$,  high purity high T$_c$ materials.
\cite{Taillefer2,Yelland,Taillefer3,Sebastian,Taillefer4,Jaudet,Boebinger}
The main frequency of oscillation of 540T in YBCO$_{6.51}$ corresponds to 2 \% of the area
of the Brillouin zone.  This area deviates dramatically from the nominal doping of 10\% expected from the Luttinger sum rule. In contrast, overdoped Tl$_2$Ba$_2$CuO$_{6+x}$ shows a single hole Fermi surface\cite{Plate} which correspond to $1+p$ holes per Cu satisfying the sum rule and further confirmed by the Hall coefficient R$_H$.\cite{Mackenzie}
A recent report displayed a  similar contrast between overdoped samples and underdoped ones in the electron-doped  Nd$_{2-x}$Ce$_x$CuO$_4$ (NCCO) where a large electron-like pocket is reported at $x=0.17$, whereas small hole-like pockets are reported at $x=0.15$ and $x=0.16$.\cite{Helm}
Shortly after the experimental discoveries, several theoretical proposals have been made.\cite{Franz,Vafek,Millis,fu-chun,Chakra,Chakra2,Jia,ChakraKee,CLMN,Podolsky,Lee,senthil}
Most of them share the underlying idea that the quantum oscillation arise due to a broken symmetry. 
\cite{Millis,fu-chun,Chakra,Chakra2,Jia,ChakraKee,CLMN,Podolsky,Lee,senthil}
Some have further offered a missing link between the Fermi arc and quantum oscillation from a closed
Fermi surface. \cite{Chakra,senthil}

However, the precise nature of the broken symmetry is still under debate.\cite{Millis,fu-chun,Chakra,Chakra2,Jia,ChakraKee,Podolsky,Lee,senthil}  Can we identify the order by analyzing the current experimental data of quantum oscillations? One salient aspect of the available experimental data is a discrepancy between experimental reports in YBCO$_{6.51}$. While they all agree that the main frequency of oscillation is about 540 T, the satellite frequencies are different.  LeBoeuf {\em et al.}\cite{Taillefer2} confirmed the first observed main frequency of 540 T.\cite{Taillefer} In addition, they found satellite frequencies of 450 T, 630 T (suggested to be due to a combination of warping of the Fermi surface and bilayer splitting), and 1130 T (suggested to be a second harmonic of the 540T main frequency). On the other hand, whereas Sebastian {\em et al.}\cite{Sebastian} found the main frequency of 540 T, they also observed an additional satellite frequency of 1650 T. Recently, Riggs {\em et al.} reported similar oscillation frequencies in specific heat measurements as those found by Sebastian {\em et al.}\cite{Boebinger} We note that multiple frequencies have so far been found only in ortho-II material and its understanding requires a study of magnetic breakdown.

The ortho-II potential present in YBCO$_{6.51}$ is not a strong potential, yet it is visible in some experimental probes. For example, while the effect of the ortho-II potential on ARPES is negligible\cite{Arpes}, the potential has been detected in Raman spectroscopy\cite{raman}.
One may naively expect that such a potential can be screened as pointed out in Ref. \onlinecite{Chakra}.
However,  first principle studies show that the oxygen ordering leads to order in the Ba position \cite{deFontaine}, which leads to a stronger ortho-II potential than one would otherwise expect.
A Fermi-surface-induced lattice modulation was observed only in oxygen empty CuO chains via diffusive x-ray scattering measurements \cite{Liu} indicating the important role of oxygen ordering in electronic structures.
It was recently suggested that the ortho-II potential can be used to
differentiate between two $(\pi,\pi)$ orders, antiferromagnetism and d-density wave orders.\cite{Podolsky}
In this scenario, the ortho-II potential plays an important role, but its relative  weakness implies that magnetic breakdown effects must be taken into account explicitly.  For instance, if the magnetic breakdown occurs below the field at which oscillations begin to appear (30 T), then the ortho-II potential can be neglected altogether in the analysis of quantum oscillations.  On the other hand, if the breakdown occurs within the currently accessible field range of 30-60T, then two different sets of Fermi surfaces -- corresponding to the Fermi surface before and after the breakdown -- may participate in the quantum oscillation.  In this case, the dominant oscillation frequencies would depend sensitively on the field range at which a particular experiment is carried out.

In this paper, we offer a general discussion on quantum oscillations in Sec. \ref{sec:oscillations} and define a way to take magnetic breakdown into account for non-trivial Fermi surface geometries. In Sec. \ref{sec:commensurate}, we focus on the AF+ortho-II system for its relevance to the ortho-II materials studied experimentally. In Sec. \ref{sec:incommensurate}, we study incommensurate orders, and show the Fermi surface of spin spiral order in the presence of ortho-II potential. We determine oscillation frequencies and magnetic breakdown fields, and compare the result with those obtained for the commensurate order. We summarize our results and discuss the implications of our findings in Sec. \ref{sec:summary}.

\section{Magnetic breakdown and Fermi surface curvature}
\label{sec:oscillations}


Quantum oscillations are a quantum mechanical effect observed in metals with closed Fermi surfaces at low temperatures in the presence of a  magnetic field.  As the intensity of the applied field increases, there are oscillations in the physical properties of the metal, including the magnetic moment (named de Haas-van Alphen, dHvA), resistivity (Shubnikov-de Haas, SdH), specific heat, and sound attenuation.  When plotted versus the inverse field, these quantities display a remarkable regularity in their oscillations.  

Quantum oscillations can be understood via the semiclassical quantization of quasiparticle energies in an applied magnetic field.  They are of great significance for their relation with Fermi surface geometry: the frequency of oscillations (in Tesla) is related to the area enclosed by a Fermi pocket (in 2 dimensions) or the area enclosed by an extremal orbit of the Fermi surface (in 3 dimensions) by the following equation:
\begin{equation}
F = \frac{\hbar c}{2\pi e}A_k ,
\label{Area}
\end{equation}
where $A_k$ is the area enclosed by the Fermi pocket and $F$ is the frequency of oscillations. If there are multiple pockets (or multiple extremal orbits), that will result in multiple frequencies that can be determined by Fourier transform.

The different frequencies are constrained by the Luttinger sum rule which relates the density of carriers to the area enclosed by the Fermi pocket through: 
\begin{equation}
p = 2 A_k \frac{ab}{(2\pi)^2},
\label{DensArea}
\end{equation}
where $a,b$ are the lattice constants and the factor of $2$ comes from spin degree of freedom. Therefore, we can relate the carrier density to the frequency of oscillation found through:
\begin{equation}
F_i = \frac{\Phi_0}{2ab}p_i,
\label{FreqDens}
\end{equation}
where $\Phi_0$ is the unit quantum flux. One must be careful when working with a broken translational symmetry to work in the reduced Brillouin zone (RBZ). In the cases that we will study, this will be important. The Luttinger sum rule specifies that the total doping $p$ is equal to the sum of carrier density for each pocket, that is
\begin{equation}
p = \sum_{i=h.l.p}p_i - \sum_{j=e.l.p}p_j ,
\label{LuttDens}
\end{equation}
where h.l.p mean hole-like pocket and e.l.p means electron-like pocket. We can easily convert this equation using the frequency of each pocket:
\begin{equation}
\frac{\Phi_0}{2ab}p = \sum_{i=h.l.p}F_i - \sum_{j=e.l.p}F_j.
\label{LuttFreq}
\end{equation}
This equation establishes a constraint on the frequencies of the system.


Magnetic breakdown is an effect that occurs at large magnetic fields for systems with multiple Fermi pockets that almost intersect one-another.
As first discussed by Cohen and Falikov\cite{CF}, magnetic breakdown happens when the transition amplitude for the quasiparticle to tunnel from one band to the other become sizeable as the magnetic field increases. A semiclassical treatment shows that magnetic breakdown occurs approximately when the condition $\hbar \omega_c \epsilon_F > E_g^2$ is satisfied\cite{Blount}. Here, $\epsilon_F$ is the Fermi energy, and $E_g$ is the energy splitting of two bands generated by a small perturbation such as a spin density wave, a charge density wave order or some lattice potential. In terms of the breakdown field ($B^*$), this can be written as
\begin{equation}
B^* = \frac{E_g^2}{v_F^2} * \frac{8\pi^2 c}{\hbar e}.
\end{equation}

A more general treatment\cite{Chambers, Shoenberg} yields a useful formula involving only the local geometry of the Fermi surface to find the breakdown field which, in the case of almost free electrons reduces to
\begin{equation}
 \label{Chambers2}
 B^* = \frac{\pi \hbar c}{e} \delta_k^2,
\end{equation}
where $\delta_k$ is the k-space distance between the two bands at the chemical potential in units of $1/a$.
This equation can also be written as
\begin{equation}
 \label{Chambers3}
 \delta_k \ell_B = const,
\end{equation}
where $\ell_B=\sqrt{\Phi_0/B^*}$. The constant of $\sqrt{2}$ in Refs. \cite{Chambers,Shoenberg} is obtained using a circular orbit as the local geometry of the Fermi surface. However in general, this constant will depend on the details of the geometry of the Fermi surface at the band separation. We therefore define a constant $K$ as
\begin{equation}
\delta_k  \ell_B = K.
\label{eq:Kdef}
\end{equation}
Below, we will show how $K$ depends on the FS curvature.

It is important to note that magnetic breakdown does not occur at a sharply-defined field, but is a crossover effect. It is usually defined to be the field at which the probability for tunnelling is equal to $exp(-1)$. In our numerical simulations below, we will define the magnetic breakdown field to be the field at which the intensity of oscillations arising from the pockets in the original Fermi surface is equal to the intensity of the pockets after reconstruction of the Fermi surface. Therefore, one should note that the intensity of oscillations from the reconstructed Fermi surface is still visible after the breakdown field as defined in this paper.

To study the effect of curvature on magnetic breakdown, we compute the value of $K$ for various systems that have different curvature. Our numerical simulations consist of solving a lattice model in an applied magnetic field.  The mean-field Hamiltonian is given by
\begin{align}
H &=& -&\sum_{ij\sigma}t_{ij}e^{iA_{ij}}c^\dagger_{i\sigma}c_{j\sigma} + \sum_i (-\mu+\lambda (-1)^{i_x}) c_{i\sigma}^\dagger c_{i\sigma}& \nonumber \\
  &&  +&\Delta_{0}\sum_{i} (-1)^i \left(c^\dagger_{i\uparrow}c_{i\uparrow}-c^\dagger_{i\downarrow}c_{i\downarrow}\right).& 
\label{tightbinding}
\end{align}
Here, $c_{i\sigma}^\dagger$ creates an electron with spin $\sigma$ at site $i$, $\mu$ is the chemical potential, $\lambda$ is the ortho-II potential, which is staggered in the $x$-direction, and $\Delta_0$ is the order parameter of the broken symmetry, like an antiferromagnet.  The amplitudes $t_{ij}$ are real, and we keep nearest neighbor ($t$), next-nearest neighbor ($t'$), and third nearest neighbor ($t''$) hoppings.  In addition, the orbital effect of magnetic field has been introduced in the standard manner through the Peierls substitution $t_{ij}\to t_{ij}e^{iA_{ij}}$, where $A_{ij}=\frac{e}{\hbar c}\int_{r_i}^{r_j} d{\bf\ell} \cdot {\bf A},$ and ${\bf A}$ is the vector potential.

\begin{figure}[]
\centering
	\subfigure[]{
	\includegraphics[height=1in,width=1in,angle=-0]{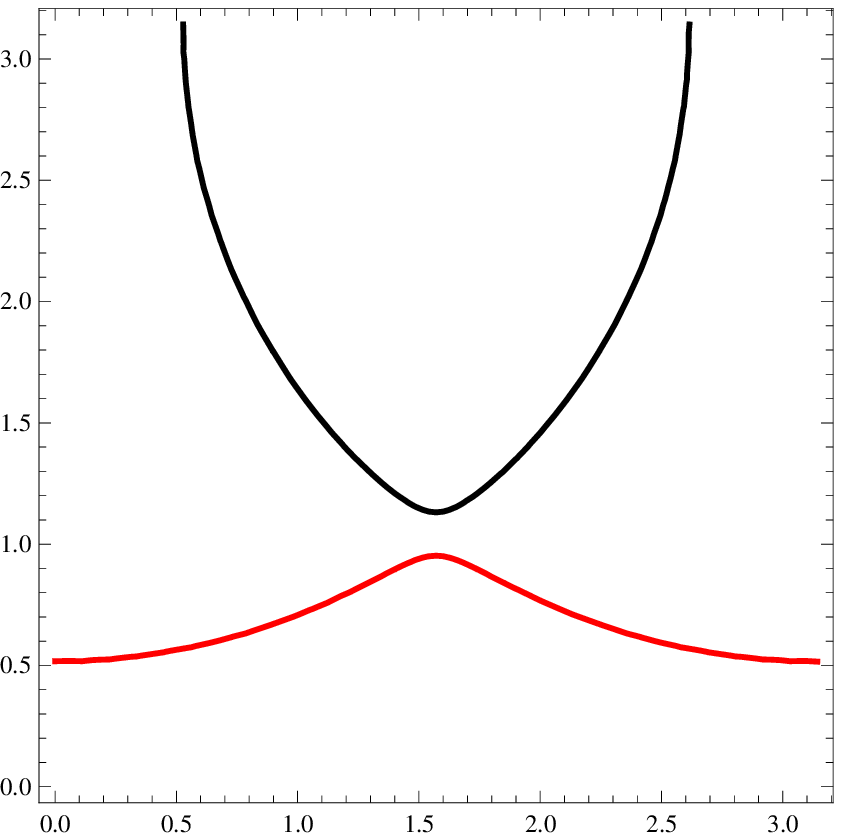}
	\label{fig:NFS:a}
	}
	\subfigure[]{
	\includegraphics[height=1in,width=1in,angle=-0]{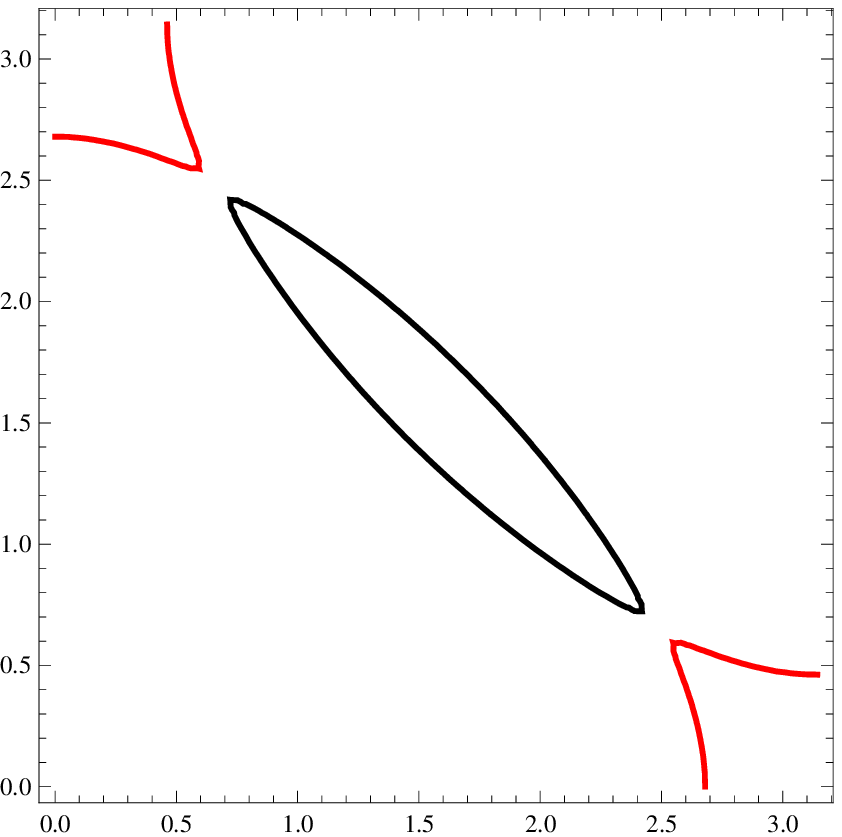}
	\label{fig:NFS:b}
	}
	\subfigure[]{
	\includegraphics[height=1in,width=1in,angle=-0]{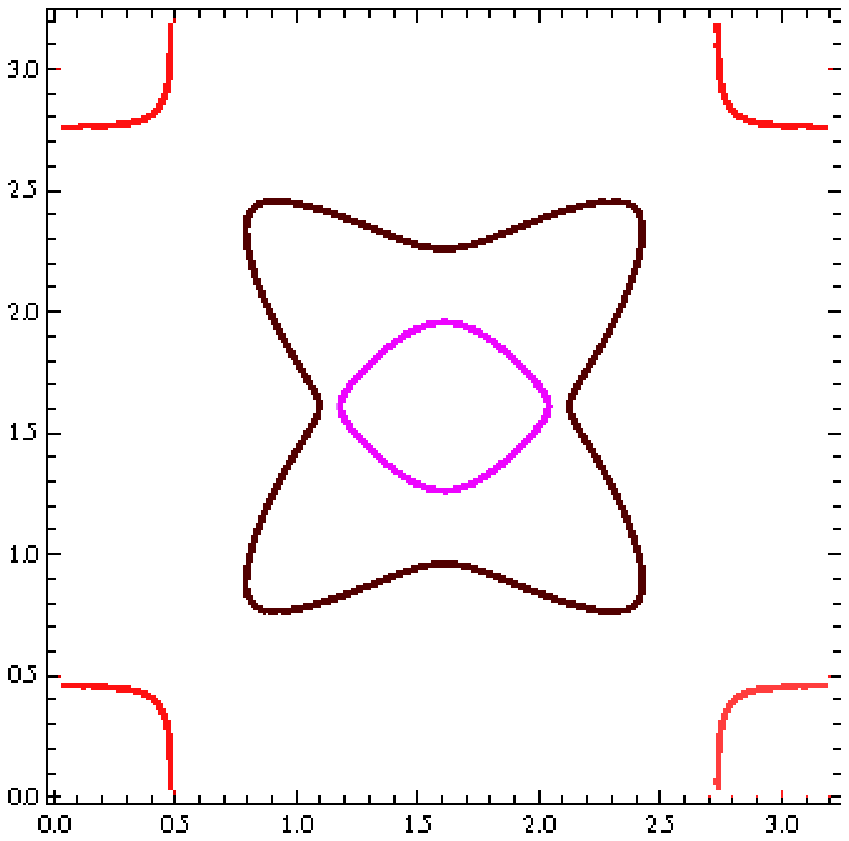}
	\label{fig:NFS:c}
	}
\caption{Fermi Surfaces for three different systems studied to explore magnetic breakdown systematically. The parameters for the electronic structure are chosen to generate different Fermi surface curvature to study its effect on magnetic breakdown. Note that the figures show only one quadrant of the Brillouin zone.}
\label{fig:NFS}
\end{figure}

The three specific systems used to study the Fermi surface curvature dependence of $K$ are (a): $\lambda$ finite and $\Delta_0 = 0$ ,  (b): $\lambda = 0$ and $\Delta_0$ finite and (c): both $\lambda$ and $\Delta_0$ finite. The Fermi surface of these three systems are displayed in Fig. \ref{fig:NFS}.

We obtain the quasiparticle spectrum of the total Hamiltonian by exact diagonalization of Bloch wave functions on a magnetic unit cell of size $L_x\times L_y$.\cite{Hofstadter}  The inclusion of an orbital magnetic field implies that, for any given fixed gauge, the vector potential appears to break translational invariance.  Translational invariance is restored to the system by noting that translations by $L_x$ and $L_y$ must be accompanied by appropriate gauge transformations.  This can only be done in a consistent fashion provided that the magnetic flux piercing the magnetic unit cell is quantized, $L_xL_y B=n\Phi_0$.\cite{Hofstadter}  Here $n$ is an integer, and $\Phi_0=hc/e$ is the quantum of flux.
In our simulations, we fix the height of the magnetic unit cell $L_y=2$, and also fix a single flux quantum per magnetic unit cell.  We then sweep the magnetic field by varying the width $L_x$ of the magnetic unit cell, $1/B=2L_x/\Phi_0$.

\begin{figure}[]
\includegraphics[height=2in,width=3in,angle=0]{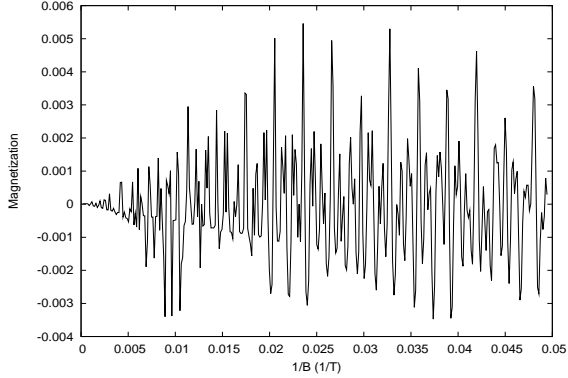}
\caption{de Haas-van Alphen effect for the AF + ortho-II simulation. The parameters used for this case, written in the $(t, t', t'', \mu, \lambda, \Delta_0)$ basis, were $(0.3, -0.09, 0.012, -0.27, 0.08, 0.07)$ }
\label{fig:dHvA}
\end{figure}

\begin{figure}
\includegraphics[height=2in,width=3in,angle=0]{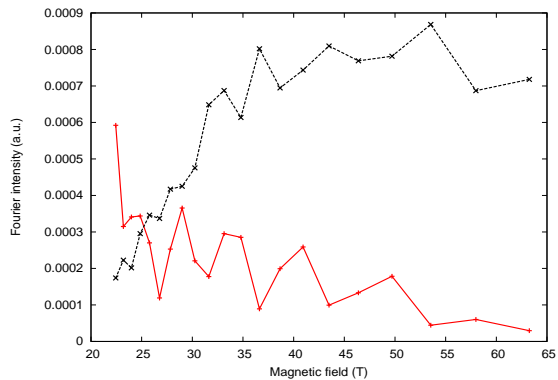}
\caption{Field-dependent Fourier intensities of the $\beta$ and $\gamma$ pockets summed together (solid line), compared to the intensity of the $\beta'$ pocket (dashed line), for the data in Fig.~\ref{fig:dHvA}.  From this graph we estimate the breakdown field $B^* \sim 25$ T.}
\label{fig:breakdown}
\end{figure}

In each system, in order to compute $K$, we vary the different parameters in order to obtain different values of $\delta_k$. For each value of $\delta_k$, we then compute the energy, the density of states (DOS) and magnetization of the system as a function of magnetic field. Fig. \ref{fig:dHvA} shows an example of magnetization as a function of the inverse field. We then compute the breakdown field by use of a ``field-dependent Fourier transform'', which consists of computing the Fourier spectrum over a narrow window of fields, and shifting the position of these windows to get a running field dependence for the Fourier spectrum.  We require the window of fields to be broad enough to contain a few oscillations of all relevant frequencies, but narrow enough to detect the magnetic breakdown. We then compare the intensities of the frequencies from the band present before the breakdown to those after the breakdown and determine the breakdown field to be the one at which the intensities are equal. Fig. \ref{fig:breakdown} shows this comparison for the example of Fig. \ref{fig:dHvA}. From this graph, we determine $B^*$ to be approximately $25T$. We compute $l_B$ from $B^*$, and obtain $\delta_k$ by determining the k-space separation between different pockets at the chemical potential; K is then simply the product of $l_B$ and $\delta_k$.

\begin{figure}
\includegraphics[height=2in,width=3in,angle=0]{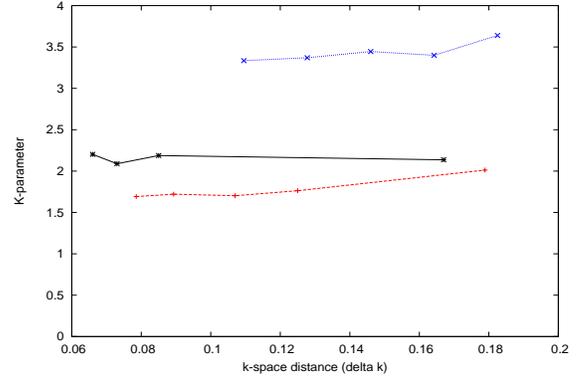}
\caption{$K$ parameter as a function of $\delta_k$. We first note the $K$ is indeed roughly constant for a given curvature. The dashed line is only ortho-II (case (a)) and the dotted line is only AF (case (b)). The solid line represents the AF + ortho-II system (case (c)). Note that $K$ is biggest for case B where the curvature of Fermi surface is sharp, so a weaker magnetic breakdown field is expected.}
\label{fig:Kparam}
\end{figure}

In system (a), shown in Fig. \ref{fig:NFS:a}, the parameters used, written in the $(t, t', t'', \mu, \lambda, \Delta_0)$ basis, were $(0.3,-0.09,0.012,-0.25,\lambda,0)$. We tuned $\delta_k$ by varying $\lambda$ over the values $\lambda = 0.022,$  $0.025,$ $0.03,$ $0.035,$ and $0.05$. The resulting value of K is shown as the dashed line in Fig. \ref{fig:Kparam} and is approximately 1.8.
For system (b), shown in Fig. \ref{fig:NFS:b}, we chose the parameters $(0.3,-0.06,0,-0.5,0,\Delta_0)$ and we tuned $\delta_k$ by varying the value of $\Delta_{0}$ over the values $\Delta_{0}=0.03,$ $0.035,$ and $0.04$. The resulting value of K is shown as the dotted line in Fig. \ref{fig:Kparam} and is approximately 3.5.
Finally for system (c), shown in Fig. \ref{fig:NFS:c}, the parameters used, written in the $(t, t', t'', \mu, \lambda, \Delta_0)$ basis, were ($0.3$, $-0.09$, $0.012$, $-0.27$, $0.08$, $0.07$), ($0.3$, $-0.095$, $0.008$, $-0.271$, $0.07$, $0.08$), ($0.3$, $-0.09$, $0.012$, $-0.27$, $0.1$, $0.07$) and ($0.3$, $-0.09$, $0.01$, $-0.27$, $0.05$, $0.21$), giving four different values of $\delta_k$.
The resulting value of K is shown as the solid line in Fig. \ref{fig:Kparam} and is approximately 2.2.
Our results confirm that B$^*$ strongly depends on the Fermi surface curvature. Note that when K is doubled, B$^*$ is quadrupled.

\section{Quantum oscillations in ortho-II + AF system}
\label{sec:commensurate}

\begin{figure}[]
\includegraphics[height=1in,width=3in,angle=-0]{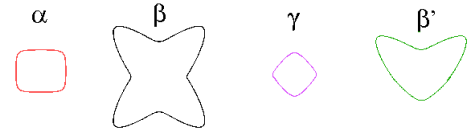}
\caption{The different Fermi pockets of the AF+ortho-II and DDW+ortho-II potentials. In a single layer system, the AF+ortho-II surface has $\alpha$, $\beta$ and $\gamma$ pockets whereas the DDW+ortho-II surface has $\alpha$ and $\beta '$ pockets.}
\label{fig:pockets}
\end{figure}

\begin{figure}
\includegraphics[height=2in,width=3in,angle=0]{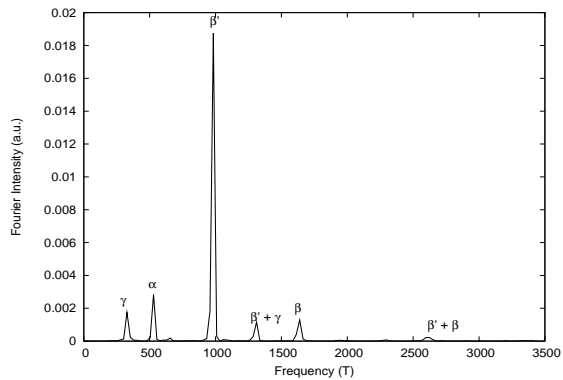}
\caption{Fourier transform of the dHvA data shown in Fig.~\ref{fig:dHvA}. This spectrum shows the following peaks: $\alpha, \beta, \gamma, \beta ', \beta ' + \gamma, \beta ' + \beta$. The parameters are $(t,t',t'',\mu,\lambda,\Delta_0) = (0.3,-0.09,0.012,-0.27,0.08,0.07)$. Here we used a Lorentzian broadening of $10^{-3}$ for Landau levels, and the field range for the transform was $20$ - $100$ T}
\label{fig:FT:a}
\end{figure}

From the three systems studied in the previous section, case (c) is the one that is relevant to the ortho-II YBCO materials. The different Fermi pockets for this system are shown and labelled in Fig. \ref{fig:pockets}. The $K$ parameter for this system was found to be around $2.2$, which differs from $\sqrt{2}$ for the circular Fermi surface \cite{Chambers, Shoenberg}. Since $B^* \propto K^2$, this difference leads to a breakdown field that is two times larger than what one would get from a naive magnetic breakdown analysis. For the parameters used in last section with $\Delta_0 = 0.07$, we expect oscillations at $F_\alpha\sim 524$ T (electron pocket), $F_\beta\sim 1630$ T (hole pocket) and $F_\gamma\sim 328$ T (hole pocket).  The smallest band splitting occurs between the $\beta$ and $\gamma$ bands.  Hence, at large magnetic fields, magnetic breakdown will lead to a combination of the $\beta$ and $\gamma$ pockets into $\beta'$ pockets, with a frequency that is approximately equal to the average of the $\beta$ and $\gamma$ frequencies, $F_\beta'\sim\frac{1630+328}{2}=979$ T.
From the magnetization shown in Fig. \ref{fig:dHvA}, a Fourier transform yields the spectrum shown in Fig. \ref{fig:FT:a} where all the expected frequencies can be seen.

To compute the breakdown field in this case, we can make use of the $K$-parameter computed in the previous section. The only thing we need in order to find $B^*$ is $\delta_k$ which can be computed analytically in this case from the dispersion of the $\beta$ and $\gamma$ along the $k_y = \pi/2$ line. These dispersions are given by
\begin{equation}
E^*_{\pm} = -2t''(\cos{2k_x}-1) -\sqrt{4t^2 \cos^2{k_x}+(\Delta \pm \lambda)^2}.
\label{eq:bgdisp}
\end{equation}
Note that $\delta_k$ is equal to the difference between the values of $k_x$ at which $E^*_\pm (k_x) = \mu$. $t$, $t''$, $\Delta$, and $\lambda$ have a direct effect on $\delta_k$, while $t'$ does not. Increasing $t$ and $t'' $ lowers $\delta_k$.  It is straightforward to see the effect if one takes either $t$ or $t''$ to 0. For example, when $t''=0$, $\delta_k = \arccos(\sqrt{\mu^2-(\Delta+\lambda)^2}/ 2t) - \arccos(\sqrt{\mu^2-(\Delta-\lambda)^2}/2t)$. If the band is normalized and the effective band mass is bigger (effectively lowering $t$), it will push B$^*$ higher. On the other hand, increasing $\Delta$ and $\lambda$ increases $\delta_k$, as we expect. While $t'$ does not have a direct effect, it affects $\delta_k$ via shifting of the chemical potential to adjust for a constant density.

While we obtain B$^* \sim 25T$ for the set of parameters given above using $K=2.2$, we emphasize that $\delta_k$ is very sensitive to the choice of parameters discussed above. For example,  $B^* \sim 100T$ for $\Delta = 0.1$ and $\lambda = 0.1$ ($\delta_k  \sim  0.13$). Eq. (\ref{eq:bgdisp}) is only true for antiferromagnetism.
Note that an ortho-II system with DDW ordering has a similar zone folding to the AF+Ortho-II case considered here.  However, as discussed in Ref.~\onlinecite{Podolsky}, in the DDW case, symmetries prevent the splitting of the $\beta$ and $\gamma$ bands. Hence, for DDW order in the presence of ortho-II potential (DDW+ortho-II order)  in monolayer systems, there are only $\alpha$ and $\beta'$ pockets and no magnetic breakdown occurs within the accessible field range.  On the other hand, in bilayer systems such as ortho-II YBCO, the $\beta$ and $\gamma$ bands do split if there is a current circulating between layers as shown in Ref.~\onlinecite{Podolsky}.  Thus, simulations of AF+ortho-II on a monolayer give us direct information regarding magnetic breakdown of DDW+ortho-II order on a bilayer, without the additional computing resources necessary to simulate the bilayer system.

\subsection{Disorder effect and anisotropic scattering}
\label{sec:disorder}
The different ortho-II materials studied so far possess a rather large mean-free path of order $l \approx 160$ \AA{}\cite{Jaudet}. Using an average Fermi velocity of $8.4$ x $10^4$ m$/$s\cite{Jaudet}, we get a scattering rate $1/\tau \approx v_F/l \approx 3$ meV.
In order to simulate the effects of disorder that exist in real materials, we broaden the Landau levels from delta functions in energy to Lorentzians of width $\Gamma\approx 1/\tau$. For instance, in the previous section, we chose $\Gamma = 1$ meV\cite{Units}, but $\Gamma$ may differ between samples. Here we study how varying the value of $\Gamma$ affects the Fourier spectrum. Table \ref{table:nonlin} shows the Fourier intensities, computed in the range of $20-70$ T, of all the pockets for different broadening. Note that the oscillation intensity is more sensitive to disorder for some pockets than for others; hence, broadening has a direct effect on B*, which is also shown in the table.

\begin{table}[ht]
\caption{Fourier spectrum as a function of broadening.}
\centering
\begin{tabular}{c|c c c c|c}
Broadening ($\Gamma$) &  \multicolumn{4}{c}{Fourier intensities (a.u.)} & B$^*$ \\
 (meV) & $I_\alpha$ & $I_\beta$ & $I_\gamma$ & $I_{\beta '}$ & (T) \\
\hline
0.6  & 50   & 100 & 20  & 190 & 25 \\
0.75 & 25   & 40  & 15  & 140 & 22 \\
1    & 10   & 10  & 15  & 80  & 23 \\
2    & 0.4  & 0.1 & 3.5 & 6   & 36 \\
3    & 0.01 & --  & 0.6 & 0.3 & 50 \\
\end{tabular}
\label{table:nonlin}
\end{table}

Note that for $\Gamma = 2$ meV, the oscillation intensity from the $\beta$ pocket, I$_\beta$ is still visible, but that for $\Gamma = 3$ meV, I$_\beta$ is gone. Hence a small difference in $\Gamma$ can make a qualitative difference in the Fourier spectrum. This is understandable because the broadening of the Landau levels will affect the larger frequencies more than the smaller ones. This is why I$_\gamma$ goes from being the smallest at $\Gamma = 0.6$ meV  to being the largest at $\Gamma = 3$ meV (see Table \ref{table:nonlin}). The $\gamma$ pocket, having the smallest frequency, is also the least affected by broadening.

Another important aspect of Fourier spectrum produced from our analysis is the difference between the relative intensities of our multiple peaks and those observed in experiment. For example, in our case, the $\beta '$ pocket is the dominant peak unless the disorder broadening is greater than 3 meV,  while its measured amplitude is small, and not confirmed by all of the experimental groups.  One way to reconcile the difference is to introduce an anisotropic scattering on the Fermi surface. If the quasiparticle scattering rate is momentum-dependent, then it is possible to obtain the electron pocket, $\alpha$, to be the dominant frequency. For example, if quasiparticles near $k=(\pi/2,\pi/2)$ (where the hole pockets are) have a larger scattering rate than the quasiparticles near $k=(\pi,0)$ and $k=(0,\pi)$ (where the electron pockets are), then the $\alpha$ pocket could get the largest intensity.  However, this contrasts with the anisotropic scattering observed by ARPES at high temperatures in the absence of a magnetic field. It was shown that the scattering rate deduced from the imaginary part of the self energy is lower along the node compared to the antinode direction\cite{Campuzano}. However, as we discussed in the introduction, the ARPES have shown the Fermi arc at high temperatures (above T$_c$) without the magnetic field,  and it is possible that the low temperature and high field state may not be smoothly connected to a high temperature and zero field pseudogap phase.

\section{Incommensurate orders}
\label{sec:incommensurate}

In both La$_{2-x}$Sr$_x$CuO$_4$ and YBa$_2$Cu$_3$O$_{6+x}$, neutron scattering experiments have reported strong inelastic signals at incommensurate wave vectors, ($\pi-\delta, \pi$).\cite{Tranquada,Stock}  Later, it was found that the quasi-elastic incommensurate peak intensity increases linearly with magnetic field in YBCO$_{6.45}$.\cite{Haug} It was also proposed that an incommensurate spiral order is induced by a magnetic field leading to a Fermi surface reconstruction responsible for the quantum oscillations.\cite{Sebastian} It is worthwhile to investigate the effect of the ortho-II potential for the incommensurate spiral order case.  While our analysis is specific to the spiral spin density wave, it can be generalized to a collinear incommensurate case if the higher order gaps are smaller than the gap generated by the ortho-II potential.\cite{Over60,Over62,Over64,Daemen} In other words, if we keep only the first order gap generated by an incommensurate collinear order our analysis leads to the same results as for the spiral case. An incommensurate spiral order breaks translational symmetry just as collinear incommensurate order does, but it mixes $\ket{k ,\uparrow}$  with $\ket{k+Q, \downarrow}$  and $\ket{k,\downarrow}$ with $\ket{k-Q, \uparrow}$. The mean field interaction term can be written as Eq. (\ref{eq:chiral}). 

\begin{equation}
H_{\rm spiral}=\Delta_{\rm sp}\sum_{\bf k} c_{{\bf k}\uparrow}^\dagger c_{{\bf k}+{\bf Q}\downarrow}+h.c.
\label{eq:chiral}
\end{equation}

\begin{figure}
\includegraphics[width=1.8in]{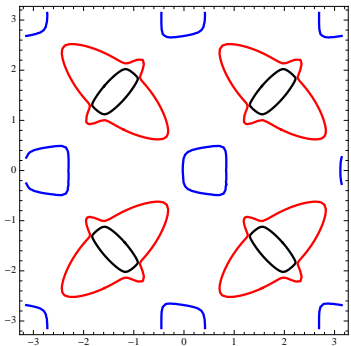}
\caption{Fermi surface for spiral spin density wave order at wave vector ${\bf Q}=(7\pi/8,\pi)$ in an ortho-II potential.  In addition, the spiral state, described by Eq. (\ref{eq:chiral}), gives rise to a second set of Fermi surfaces, not shown here, obtained by setting ${\bf Q}\to-{\bf Q}$ and flipping up and down spins.}
\label{fig:chiralFS}
\end{figure}

The Fermi surface for the choice of parameters $t=0.3$, $t'=-0.09$, $t''=0.012$, $\mu=-0.27$, $\lambda=0.05$, and $\Delta_{\rm sp}=0.08$, and for the wave vector ${\bf Q}=(7\pi/8,\pi)$ is shown in Fig. \ref{fig:chiralFS}.

\begin{figure}
\includegraphics[width=2.6in]{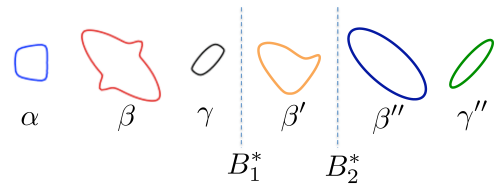}
\caption{Pockets arising from the Fermi surface in Fig.~\ref{fig:chiralFS}.  Oscillations at the electron hole frequency $F_\alpha\sim 525$ T should be seen at all experimentally-relevant values of field.  On the other hand, the hole pockets undergo breakdown at two different values of field.  For $B<B_1^*,$ oscillations occur at $F_\beta\sim 1650$ T and $F_\gamma\sim 278$ T.  At intermediate fields, $B_1^*<B<B_2^*,$ hole pocket oscillations are dominated by the frequency $F_{\beta'}\sim 965$ T.  For $B>B_2^*$, oscillations occur at $F_{\beta''}\sim 1505$ T and $F_{\gamma''}\sim 425 $ T. Based on the value of $\delta_k$ responsible for the two breakdowns, we expect $B_2^*/B_1^*\sim 5$.}
\label{fig:chiralpockets} 
\end{figure}

From Fig. \ref{fig:chiralFS}, there are now two Fermi surface splittings between $\beta$ and the $\gamma$ pockets which may be relevant to magnetic breakdown. We can determine the two different band separation at the chemical potential ($\delta_k$) that will then give us an estimate for the two breakdown field $B_1^*$ and $B_2^*$ (see fig. \ref{fig:chiralpockets}. They are respectively $\delta_{k1} = 0.07$ and $\delta_{k2} = 0.19$. Assuming that K is of order 2 (a reasonable estimate since the curvatures at the two separations are very similar to the  commensurate case), then we get the following values for the two breakdown fields: $B_1^* = 35T$ and $B_2^* = 251T$.

\section{Summary and Discussion}
\label{sec:summary}

ARPES and dHvA/SdH oscillations are two of the most powerful tools to measure the Fermi surface of a metal.  The two methods give complimentary information about a system. Quantum oscillation experiments probe the bulk of a system, and give a measurement of the size of various Fermi pockets in the system.  However, they reveal neither the location of the pockets, nor the number of different pockets of equal area.  In addition, quantum oscillation experiments must be carried out in a magnetic field.  In contrast, ARPES, which can give details of the locations and shapes of different pockets in the Brillouin zone, is a surface probe which must be carried out in the absence of a magnetic field. Therefore, a combination of the two techniques should offer a relatively complete picture of the Fermi surface of a material.

In the high T$_c$ cuprates, there is a qualitative discrepancy between the results provided by the two measurements:
ARPES measurements indicate Fermi arcs, while quantum oscillations results reveal the existence of closed Fermi surfaces.
Several theoretical\cite{Millis, ChakraKee} and experimental proposals  support an electron pocket based on the negative Hall coefficient measured at low temperatures\cite{Taillefer2} and more recently, high-field thermoelectric measurements performed on a series of high T$_c$ materials also indicate that this 540T frequency comes from an electron-like pocket\cite{Chang}. However, there is no visible spectral weight such as an electron pocket near ($\pi,0$) in ARPES. There is also a discrepancy in the presence of satellite frequencies, such as 1650T, among the different quantum oscillation experiments and related proposals.

Another important experimental observation is the absence of a visible Zeeman effect on the phase of oscillation.\cite{Sebastian2} Sebastian {\em et al} proposed that the order responsible for quantum oscillations  is in the spin triplet channel. While the DDW discussed above is in the singlet channel, it is interesting to note that nematicity allows a coupling between spin triplet DDW and AF order in the presence of a magnetic field.\cite{Kee} The nematicity, a broken x-y symmetry is reported in YBCO by neutron scattering.\cite{Keimer} Such a coupling leads to a magnetic field-induced ordering which is a combination of AF and triplet DDW, where the dominant order is the one with the smaller gap in the spectrum\cite{Kee} 

In this paper, motivated by a series of quantum oscillation experiments performed on ortho-II YBCO$_{6.51}$ at low temperatures
and high magnetic fields, we investigated the magnetic breakdown effect and the  Fermi surface topology in ortho-II high T$_c$ cuprates and showed how to reconcile the discrepancy in the observed satellite frequencies. Our results also pose constraints in a theory of quantum oscillations in high T$_c$ cuprates in general. We provided a general criterion for the magnetic breakdown, showing that it depends, in addition to a band separation and the Fermi velocity derived from the semiclassical study, on the Fermi surface curvature.

Applying our results to ($\pi,\pi$) orders such as AF in the presence of ortho-II potential, we found that the magnetic breakdown field
is highly sensitive to the ortho-II potential,  the electronic dispersion, and the order parameter strength.
As expected, the magnetic breakdown field increases when the potential increases, the AF order get stronger, or the electronic dispersion gets flatter. Smoother Fermi surface curvature, modified by the potential, also increases the breakdown field.
We found that the magnetic breakdown field (defined as the field strength at which the intensities of the original and modified frequencies match) can change from 25 T to 100 T.  For example, a factor of 4 increase can be achieved if the ortho-II potential ($\lambda$) increases from 0.08 to 0.1 and the AF order strength ($\Delta_{}$) from 0.07 to 0.1. DDW in bilayers gives similar results, if there is a circulating current between the layers.

While the shape of the Fermi surface depends on an exact broken symmetry (as shown in Fig. 2 for the AF + ortho-II case and in Fig. 10 for the incommensurate spiral spin density wave), the presence of multiple frequencies strongly supports a  broken translational symmetry.
Among several frequencies observed in experiments, the presence of the 1650T frequency can be explained by a $\beta$-pocket that occurs due to a near ($\pi,\pi$)-folding and a ($\pi,0$)-folding (from ortho-II potential) of the Fermi surface and that such pocket stop yielding oscillations beyond B$^*$.

In our analysis,  the intensity of each frequency depends on the disorder strength.
A disorder broadening $\Gamma$ bigger than 3 meV (which is consistent with experimental values)  makes the $\beta$ pocket, corresponding to the 1650T frequency, disappear.
This is simply because  a larger broadening makes faster oscillations (larger frequency) less visible.
The disorder strength also affects the intensity of the other frequencies. 
As demonstrated in Table I,  the $\gamma$ pocket has the highest intensity for a  3 meV broadening.
One could argue that the main frequency is a hole pocket (like the $\gamma$ pocket in our model)\cite{gamma}.
However, the difficulty with that argument is that a main frequency of 660 T was observed in YBa$_2$Cu$_4$O$_8$, an ortho-II free material.
As pointed out in Ref. \onlinecite{Podolsky}, the electron pocket is insensitive to the ortho-II potential while the hole pocket
is strongly modified. Therefore if a similar small frequency is a universal phenomenon in high T$_c$ cuprates,  the main frequency should be one that is not modified by the ortho-II potential, such as the $\alpha$-pocket in this model.

In addition, we extended our study to systems with incommensurate orders, which undergo a series of breakdowns with distinctly different Fermi surface shapes. To make progress in understanding the topology of the Fermi surface and associated order, a systematic study of the magnetic field angle dependence on the quantum oscillations and further ARPES studies in both ortho-II and ortho-II-free materials are necessary.

\textit{Acknowledgements} We are grateful to N. Harrison, S. Kivelson, S. Sebastian, and  L. Taillefer for useful discussions. This work was supported by NSERC of Canada, Canadian Institute for Advanced Research, and Canada Research Chair. This work was also supported in part by the NSF under the grant PHY05-51164 at the KTTP.



\begin{thebibliography}{9}
\bibitem{Units} The unit of meV is used assuming $t = 0.3 \equiv 300$ meV.
\bibitem{gamma} While the current choice of parameter leads to F$_\alpha = 524$ T, F$_\beta = 1630$ T and F$_\gamma = 328$ T, the frequencies of oscillations are sensitive to the choice of parameters as discussed in Ref. \onlinecite{Podolsky}. The only constraint in the frequencies
is that they obey the sum rule as shown in Eq. \ref{LuttFreq} for a doping of 10$\%$. Therefore  it is possible that F$_\gamma \sim 500$T, and F$_\alpha \sim 600-700$T, leading to F$_\beta \sim 1500-1600$T and F$_\beta' \sim 1000-1050$ T.
\end{thebibliography}
\end{document}